\documentclass{emulateapj}

\usepackage{amsmath,amssymb}
\usepackage{graphicx}

\newcommand{\um}{\mu\textnormal{m}}
\newcommand{\Rp}{R_p}
\newcommand{\Rs}{R_{\star}}
\newcommand{\RpRs}{\Rp/\Rs}
\newcommand{\aRs}{a/\Rs}

\newcommand{\Tmid}{T_{\textnormal{mid}}}
\newcommand{\Teq}{T_{\textnormal{eq}}}
\newcommand{\MJ}{M_{\textnormal{J}}}
\newcommand{\RJ}{R_{\textnormal{J}}}
\newcommand{\watersig}{5.4}

\begin{document}

\title{Detection of H2O and evidence for TiO/VO in an ultra hot exoplanet atmosphere}

\author{Thomas~M.~Evans\altaffilmark{1,$\dagger$}, David K.\ Sing\altaffilmark{1}, H.\ R.\ Wakeford\altaffilmark{2}, N.\ Nikolov\altaffilmark{1}, G.\ E.\ Ballester\altaffilmark{3}, B.\ Drummond\altaffilmark{1}, T.\ Kataria\altaffilmark{1}, N.\ P.\ Gibson\altaffilmark{4}, D.\ S.\ Amundsen\altaffilmark{5,6}, J.\ Spake\altaffilmark{1}}

\altaffiltext{1}{ School of Physics, University of Exeter, EX4 4QL Exeter, UK }
\altaffiltext{2}{ NASA Goddard Space Flight Center, Greenbelt, MD 20771, USA }
\altaffiltext{3}{ Lunar and Planetary Laboratory, University of Arizona, Tucson, Arizona 85721, USA }
\altaffiltext{4}{ Astrophysics Research Centre, School of Mathematics and Physics, Queens University Belfast, Belfast BT7 1NN, UK }
\altaffiltext{5}{ Department of Applied Physics and Applied Mathematics, Columbia University, New York, NY 10025, USA }
\altaffiltext{6}{ NASA Goddard Institute for Space Studies, New York, NY 10025, USA }

\altaffiltext{$\dagger$}{tevans@astro.ex.ac.uk}


\begin{abstract}
We present a primary transit observation for the ultra hot ($\Teq \sim 2400$\,K) gas giant expolanet WASP-121b, made using the \textit{Hubble Space Telescope} Wide Field Camera 3 in spectroscopic mode across the 1.12--1.64$\um$ wavelength range. The 1.4$\um$ water absorption band is detected at high confidence ($\watersig\sigma$) in the planetary atmosphere. We also reanalyze ground-based photometric lightcurves taken in the $B$, $r^\prime$, and $z^\prime$ filters. Significantly deeper transits are measured in these optical bandpasses relative to the near-infrared wavelengths. We conclude that scattering by high-altitude haze alone is unlikely to account for this difference, and instead interpret it as evidence for titanium oxide and vanadium oxide absorption. Enhanced opacity is also inferred across the $1.12$--$1.3\um$ wavelength range, possibly due to iron hydride absorption. If confirmed, WASP-121b will be the first exoplanet with titanium oxide, vanadium oxide, and iron hydride detected in transmission. The latter are important species in M/L dwarfs, and their presence is likely to have a significant effect on the overall physics and chemistry of the atmosphere, including the production of a strong thermal inversion.
\end{abstract}


\section{ Introduction } \label{sec:introduction}

Observations have revealed a diversity of atmospheres across the population of transiting gas giant exoplanets \citep[][]{2016Natur.529...59S,2016ApJ...817L..16S}. Transmission spectra at optical wavelengths show evidence for Rayleigh scattering by molecular hydrogen and high-altitude aerosols, and absorption by alkali metals \citep[e.g.][]{2002ApJ...568..377C, 2008MNRAS.385..109P, 2013MNRAS.436.2956S, 2015MNRAS.446.2428S, 2014MNRAS.437...46N, 2015MNRAS.447..463N}. In the near-infrared, water absorption has now been robustly measured for a number of planets, both in transmission and emission \citep[e.g.][]{2013ApJ...774...95D, 2013MNRAS.435.3481W, 2014Sci...346..838S, 2014ApJ...791...55M, 2015ApJ...814...66K}.

At temperatures above $2000$\,K, models predict that gaseous titanium oxide (TiO) and vanadium oxide (VO) are important absorbers, especially at optical wavelengths \citep[e.g.][]{2003ApJ...594.1011H,2007ApJS..168..140S,2008ApJ...678.1419F}. Indeed, TiO/VO absorption is prominent in late M and early L dwarf atmospheres at these temperatures \citep{1999ApJ...512..843B,1999ApJ...519..802K,2001RvMP...73..719B}. If TiO and VO were present in the upper atmosphere of an irradiated gas giant, it would likely generate a thermal inversion, as incoming stellar radiation is absorbed at low pressures \citep{2003ApJ...594.1011H,2008ApJ...678.1419F}.  Although thermal inversions have previously been claimed for a number of exoplanets, most of these results were based on \textit{Spitzer Space Telescope} secondary eclipse measurements since called into question \citep{2014ApJ...796...66D,2014MNRAS.444.3632H,2015MNRAS.451..680E}. Furthermore, searches for TiO/VO in the hottest gas giants have failed to detect either species in transmission \citep[][]{2013MNRAS.434.3252H,2013MNRAS.436.2956S}. 

One possibility is that TiO/VO is depleted from the upper atmosphere by cold-trapping, either deeper in the dayside atmosphere or on the cooler nightside \citep{2003ApJ...594.1011H,2009ApJ...699.1487S}. The hottest planets, however, might avoid cold-traps. \cite{2015ApJ...806..146H} have recently published near-infrared secondary eclipse observations for the gas giant WASP-33b, which has an equilibrium temperature of $\Teq \sim 2700$\,K.\footnote{We quote equilibrium temperature $\Teq$ as the blackbody temperature required for planetary thermal emission to balance the stellar irradiation, assuming zero Bond albedo and uniform day-night recirculation. Due to the approximate nature of equilibrium temperature, we round values to the nearest 100\,K in this Letter.} These data reveal excess emission around $1.2\um$, which is consistent with TiO and would imply a thermal inversion. To confirm this hypothesis, further observations are required to spectrally resolve the TiO bandheads. 

In this Letter, we present near-infrared transmission spectroscopy observations for another ultra hot gas giant, WASP-121b \citep{2016MNRAS.tmp..312D}. This extreme planet is in a 1.3 day polar orbit around an F6V host star, with a semimajor axis of 0.025\,AU. This puts WASP-121b just beyond the Roche limit, where it is subject to intense tidal forces. With an equilibrium temperature of $\Teq \sim 2400$\,K, the atmosphere may be hot enough for gaseous TiO/VO to be abundant. WASP-121b is also one of the most inflated planets known, with mass $1.2\MJ$ and radius $1.8\RJ$. These properties, combined with the brightness of the host star ($J=9.6$\,mag), make WASP-121b an excellent target for atmospheric characterization.

\section{Observations} \label{sec:observations}

A single transit of WASP-121b was observed on 2016 February 6 UT using the \textit{Hubble Space Telescope} (\textit{HST}) Wide Field Camera 3 (WFC3) for Program 14468 (P.I.\ Evans). Spectroscopic mode was used with grism G141 and forward scanning to allow longer exposures without saturating the detector \citep{2012wfc..rept....8M}. To reduce overheads, a $256 \times 256$ subarray containing the target spectrum was sampled with 15 non-destructive reads per exposure. Integration times were 103 seconds, using the SPARS10 readout mode.\footnote{See the WFC3 handbook at http://www.stsci.edu/hst/wfc3} A scan rate of 0.12 arcsec/sec was adopted, giving scans along 100 pixel rows of the cross-dispersion axis for each exposure. Typical count levels remained below $2.2 \times 10^4$ electrons, well within the linear regime of the detector.

Exposures were taken contiguously over five \textit{HST} orbits, with 17 exposures obtained per orbit. The first orbit allowed the telescope to settle into its new pointing position, and was discarded during the lightcurve analysis due to the large amplitude instrumental systematics it exhibited, as is standard practice for exoplanet transit observations \citep[e.g.][]{2013ApJ...774...95D,2013MNRAS.434.3252H,2014Sci...346..838S,2015ApJ...806..146H}. Of the remaining four \textit{HST} orbits, the first and fourth provided out-of-transit baseline, while the second and third occurred during the transit.

\section{Data reduction} \label{sec:data_reduction}

Our analysis commenced with the IMA data files produced by the CALWF3 pipeline (v3.1.6), which already have basic calibrations such as flat fielding and bias subtraction applied. We extracted flux for WASP-121 from each exposure by taking the difference between successive non-destructive reads. For each read difference, we removed the background by taking the median flux in a box of pixels well away from the stellar spectra. Typical background levels were 115 electrons. We then determined the flux-weighted center of the WASP-121 scan, and set to zero all pixel values located more than 35 pixels above and below along the cross-dispersion axis. Application of this tophat filter had the effect of masking the flux contributions from nearby contaminants, including a faint star separated by 7 arcsec on the sky. It had the additional advantage of eliminating many of the pixels affected by cosmic rays. Final reconstructed images were produced by adding together the read differences for each exposure. We extracted spectra from the reconstructed images by summing the flux within rectangular apertures centered on the scanned spectra and spanning the full dispersion axis. We experimented with aperture widths along the cross-dispersion axis ranging from 100 to 200 pixels in increments of 10 pixels, and found that an aperture of 160 pixels minimized the residuals in the final white lightcurve fit (Section \ref{sec:lightcurve_analysis}). The wavelength solution was determined by cross-correlating the first spectrum, with dispersion 4.64\,nm/pixel, against an ATLAS stellar model \citep{1993KurCD..13.....K} with properties similar to WASP-121 ($T_{\textnormal{eff}}=6500$\,K, $\log g = 4.0$\,cgs, $v_{\textnormal{turb}}=2$\,km/s) modulated by the throughput of the G141 grism. Prior to cross-correlation, both the measured spectrum and ATLAS spectrum were smoothed using a Gaussian filter with full-width half maximum of 20nm. As a result, the cross-correlation was most sensitive to the steep edges of the G141 response curve, rather than individual stellar lines. We repeated this process for all remaining spectra to determine shifts along the dispersion axis over the course of the observations, which were found to be within 0.23 pixels.

\section{Lightcurve analysis} \label{sec:lightcurve_analysis}

\begin{figure}
\centering  
\includegraphics[width=\columnwidth]{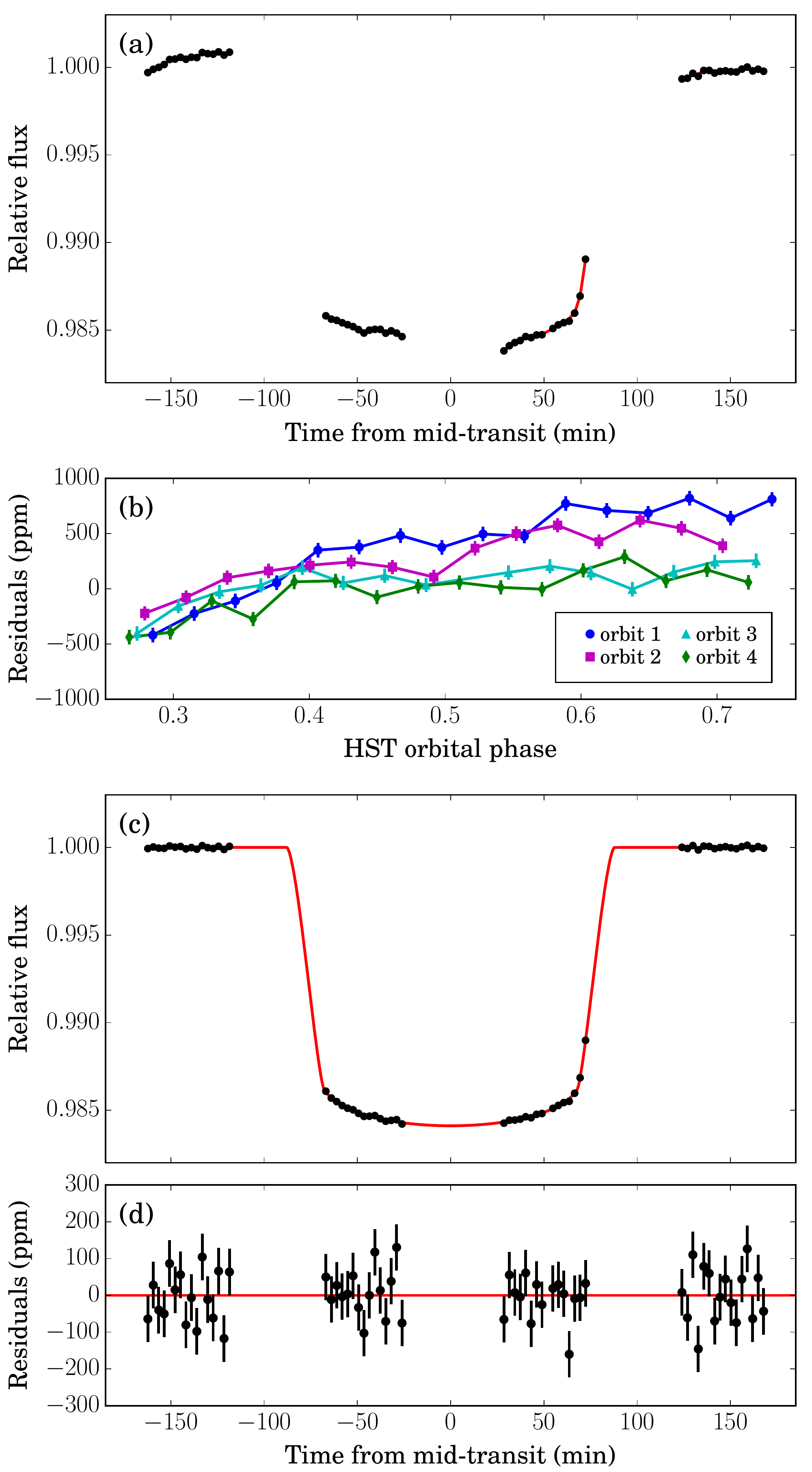}
\caption{\textit{HST}/WFC3 G141 white transit lightcurve for WASP-121b. (a) Raw flux time series, with solid lines indicating the predictive mean of the maximum likelihood GP model. (b) Residuals after removing the transit and linear time trend, which correlate with \textit{HST} phase and dispersion drift. (c) Relative flux as a function of time after removing the linear time trend and systematics component of the GP model, with solid line indicating the inferred transit signal. (d) Residuals as a function of time with photon noise errorbars after removing the combined transit, linear time trend, and GP systematics model.}
\label{fig:white_lightcurve}
\end{figure}

White lightcurves were generated for each trial aperture by summing the flux for each spectrum along the dispersion axis. We discarded the first exposure from each orbit, as these had significantly lower counts than subsequent exposures. Two additional frames were flagged as outliers, with closer inspection revealing that one was affected by a cosmic ray and the other exhibited an anomalous scan. The resulting lightcurve obtained for the 160 pixel aperture is shown in Figure \ref{fig:white_lightcurve}. Instrumental systematics that correlate with the \textit{HST} orbital phase are evident, caused by the varying thermal environment experienced by the spacecraft during its orbit \citep[e.g.][]{2013MNRAS.436.2956S,2016ApJ...819...10W}. 

\begin{table}
\begin{minipage}{\columnwidth}
  \centering
\caption{White lightcurve fit results and adopted parameters \label{table:white_lightcurve}}
\begin{tabular}{cc} 
\hline \\ 
Parameter & Value \medskip \\ 
\hline \\ 
\smallskip $\RpRs$ & ${0.12109}_{-0.00032}^{+0.00031}$ \\ \smallskip 
 $\Tmid$ (HJD$_{\textnormal{UTC}}$) & ${2457424.88307}_{-0.00011}^{+0.00010}$ \\ \smallskip 
 $\aRs$ & $3.754$ \\ \smallskip 
 $b$ & $0.160$ \\ \smallskip 
 $i \ (^{\circ})$ & $87.557$ \\ \smallskip 
 $c_1$ & $0.582$ \\ \smallskip 
 $c_2$ & $0.151$ \\ \smallskip 
 $c_3$ & $-0.435$ \\ \smallskip 
 $c_4$ & $0.199$ \\ \hline 
\end{tabular}
\end{minipage}
\end{table}

To model the white lightcurves, we adopted a \cite{2002ApJ...580L.171M} analytic function for the planet signal. For stellar limb darkening, we used the four-parameter nonlinear limb law of \cite{2000A&A...363.1081C} with coefficients $(c_1, c_2, c_3, c_4)$ listed in Table \ref{table:white_lightcurve}. The latter were obtained by fitting to the limb-darkened intensities of the ATLAS stellar model described in Section \ref{sec:data_reduction} multiplied by the G141 throughput profile.

 For the white lightcurve fitting, we treated the data as a Gaussian process (GP), using the approach described in \cite{2012MNRAS.419.2683G}. For the mean function, we adopted a \cite{2002ApJ...580L.171M} transit model multiplied by a linear time trend, and for the covariance matrix, we used a Mat\'{e}rn $\nu=3/2$ kernel \citep[see][]{2013MNRAS.428.3680G} with \textit{HST} orbital phase and dispersion shift as the input variables. We allowed the linear trend and covariance parameters to vary in the fitting, along with the planet-to-star radius ratio $\RpRs$ and transit mid-time $\Tmid$. The orbital period $P$, normalized semimajor axis $\aRs$, orbital inclination $i$, and eccentricity $e$ were fixed to values reported in \cite{2016MNRAS.tmp..312D} and listed in Table \ref{table:white_lightcurve}. Uniform priors were adopted for $\RpRs$, $\Tmid$, and the linear trend parameters. Gamma priors of the form $\textnormal{Gam}(\alpha=1,\beta=100)$ and $\textnormal{Gam}(\alpha=1,\beta=1)$ were adopted for the covariance amplitude and inverse correlation length scales, respectively, giving preference to simpler systematics models, i.e.\ smaller covariance amplitudes and longer correlation length scales. Maximum likelihood solutions were located using nonlinear optimization, as implemented by the \texttt{fmin} routine of the \texttt{scipy.optimize} Python software package.\footnote{http://scipy.org} The likelihood distribution was then marginalized using affine-invariant Markov chain Monte Carlo (MCMC), as implemented by the \texttt{emcee} software package \citep{2013PASP..125..306F}. This was done by initializing 100 walkers in close proximity to the maximum likelihood solution, and allowing them to run for 300 steps. Correlation length scales were calculated for each parameter, and a burn-in phase of three times the longest correlation length scale was discarded from all walker chains before combining them into a single chain. The resulting posterior samples displayed good mixing and convergence. Best-fit model residuals for the lightcurve generated with the 160 pixel aperture (Section \ref{sec:data_reduction}) had the lowest scatter, within 5\% of the photon noise floor. We adopt this reduction for all subsequent analysis, and report the results in Table \ref{table:white_lightcurve} with best-fit model shown in Figure \ref{fig:white_lightcurve}.

\begin{figure*}
\centering  
\includegraphics[width=\linewidth]{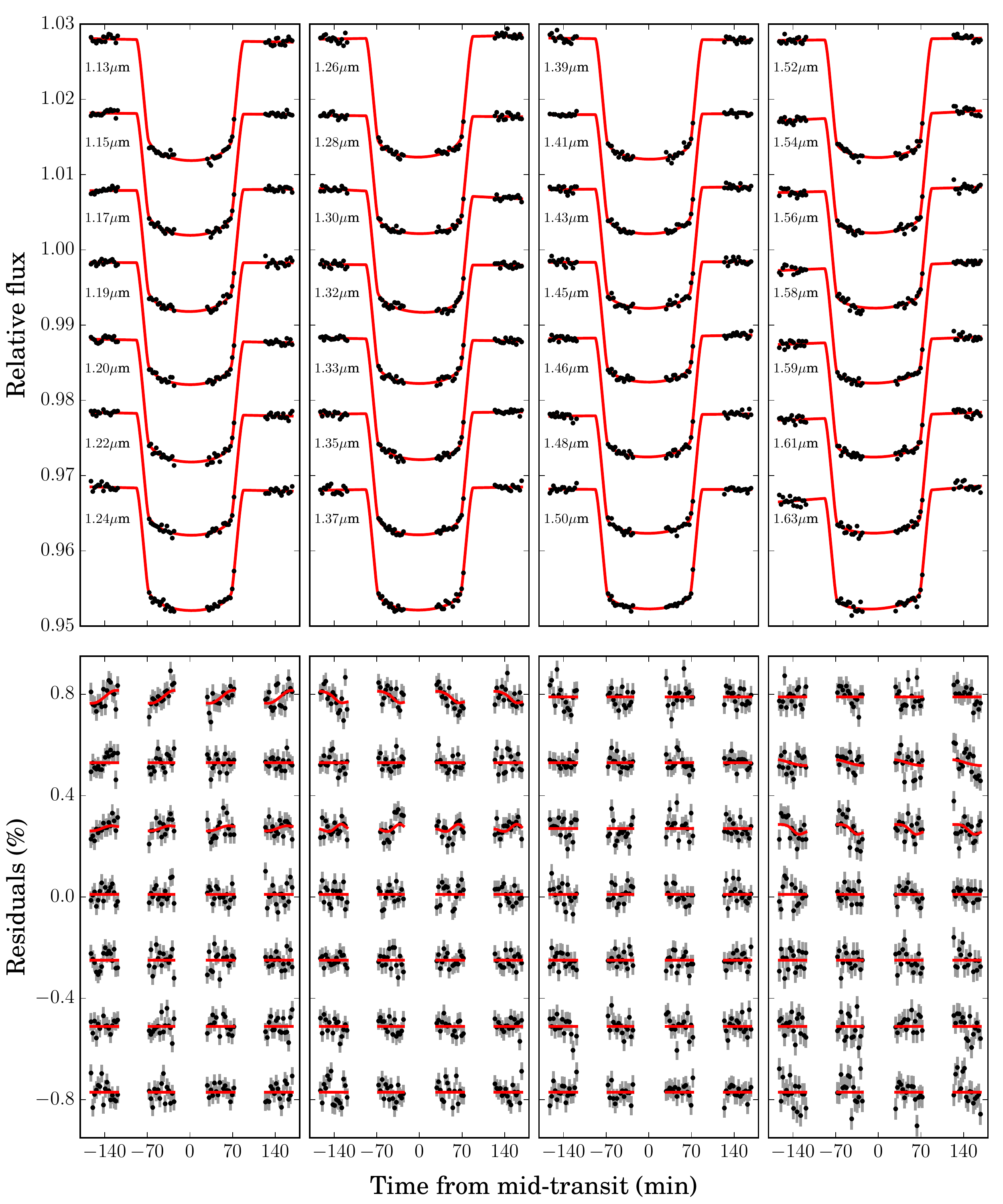}
\caption{ (Top) Raw spectroscopic lightcurves with solid lines showing best-fit transit signals multiplied by linear time trends. (Bottom) Model residuals with photon noise errorbars in gray and solid lines showing GP model fits. Note that the transit signals, linear time trends, and GP systematics models were fit simultaneously in practice, but have been separated here for illustrative purposes.}
\label{fig:spectroscopic_lightcurves}
\end{figure*}

After fitting the white lightcurve, spectroscopic lightcurves were produced using a similar approach to \cite{2013ApJ...774...95D}. First, a reference spectrum was produced by taking the average of the out-of-transit spectra. Each individual spectrum was then shifted laterally in wavelength and stretched vertically in flux to match the reference spectrum, using linear least squares. The residuals of these fits were binned into 28 spectroscopic channels across the $1.12$--$1.64\,\um$ wavelength range, each spanning 4 columns of the dispersion axis. The spectroscopic residuals were then added to a transit signal with $\RpRs$ and $\Tmid$ set to the white lightcurve best-fit values (Table \ref{table:white_lightcurve}), and limb darkening appropriate to the wavelength channel, giving the final spectroscopic lightcurves shown in Figure \ref{fig:spectroscopic_lightcurves}. This process reduces systematics that are common-mode in wavelength, as well as those arising due to the spectra drifting along the dispersion axis. It retains, however, channel-to-channel differences in the flux level, including possible transit depth variations caused by the wavelength-dependent opacity of the planetary atmosphere. 

We fit the spectroscopic lightcurves using the same GP treatment described above for the white lightcurve. For these fits, $\RpRs$ was allowed to vary separately for each channel, while $\Tmid$ was held fixed to the best-fit white lightcurve value. Again, a nonlinear limb darkening law was used with coefficients fixed to the values reported in Table \ref{table:spectroscopic_lightcurves}, which were determined using the same method described in Section \ref{sec:data_reduction} for the white lightcurve. We report the results for $\RpRs$ in Table \ref{table:spectroscopic_lightcurves} and show best-fit models in Figure \ref{fig:spectroscopic_lightcurves}. 

Following \cite{2016ApJ...819...10W}, we also performed an independent analysis of the white and spectroscopic lightcurves using the method of \cite{2014MNRAS.445.3401G} in which marginalization was performed over a grid of parametric systematics models. This gave results in good agreement with the GP analysis.

\begin{table}
\begin{minipage}{\columnwidth}
  \centering
\caption{Inferred transmission spectrum and adopted nonlinear limb darkening coefficients \label{table:spectroscopic_lightcurves}}

\begin{tabular}{cccccc} 
\hline \\ 
$\lambda$ ($\um$) & $\RpRs$ & $c_1$ & $c_2$ & $c_3$ & $c_4$ \medskip \\ 
\hline \\ 
 \smallskip 0.392--0.481 & ${0.12375}_{-0.00069}^{+0.00061}$ &  $0.337$ & $0.653$ & $-0.098$ & $-0.041$ \\ \smallskip 
0.555--0.670 & ${0.12521}_{-0.00069}^{+0.00065}$ &  $0.465$ & $0.553$ & $-0.408$ & $0.104$ \\ \smallskip 
0.836--0.943 & ${0.12298}_{-0.00117}^{+0.00114}$ &  $0.492$ & $0.309$ & $-0.311$ & $0.083$ \\ \smallskip 
1.121--1.139 & ${0.12033}_{-0.00038}^{+0.00038}$ &  $0.507$ & $0.201$ & $-0.268$ & $0.086$ \\ \smallskip 
1.139--1.158 & ${0.12107}_{-0.00036}^{+0.00035}$ &  $0.501$ & $0.233$ & $-0.318$ & $0.109$ \\ \smallskip 
1.158--1.177 & ${0.12122}_{-0.00033}^{+0.00035}$ &  $0.502$ & $0.239$ & $-0.346$ & $0.124$ \\ \smallskip 
1.177--1.195 & ${0.12149}_{-0.00034}^{+0.00035}$ &  $0.500$ & $0.255$ & $-0.377$ & $0.139$ \\ \smallskip 
1.195--1.214 & ${0.12116}_{-0.00034}^{+0.00034}$ &  $0.497$ & $0.270$ & $-0.404$ & $0.151$ \\ \smallskip 
1.214--1.232 & ${0.12113}_{-0.00036}^{+0.00033}$ &  $0.491$ & $0.319$ & $-0.482$ & $0.186$ \\ \smallskip 
1.232--1.251 & ${0.12143}_{-0.00035}^{+0.00034}$ &  $0.487$ & $0.338$ & $-0.519$ & $0.205$ \\ \smallskip 
1.251--1.269 & ${0.12070}_{-0.00036}^{+0.00037}$ &  $0.496$ & $0.325$ & $-0.518$ & $0.205$ \\ \smallskip 
1.269--1.288 & ${0.12036}_{-0.00035}^{+0.00036}$ &  $0.543$ & $0.236$ & $-0.486$ & $0.194$ \\ \smallskip 
1.288--1.307 & ${0.12033}_{-0.00037}^{+0.00036}$ &  $0.500$ & $0.343$ & $-0.565$ & $0.227$ \\ \smallskip 
1.307--1.325 & ${0.12035}_{-0.00033}^{+0.00033}$ &  $0.499$ & $0.364$ & $-0.620$ & $0.259$ \\ \smallskip 
1.325--1.344 & ${0.12118}_{-0.00034}^{+0.00034}$ &  $0.507$ & $0.356$ & $-0.628$ & $0.266$ \\ \smallskip 
1.344--1.362 & ${0.12201}_{-0.00034}^{+0.00035}$ &  $0.516$ & $0.342$ & $-0.630$ & $0.271$ \\ \smallskip 
1.362--1.381 & ${0.12193}_{-0.00035}^{+0.00034}$ &  $0.528$ & $0.314$ & $-0.609$ & $0.264$ \\ \smallskip 
1.381--1.399 & ${0.12138}_{-0.00036}^{+0.00035}$ &  $0.534$ & $0.327$ & $-0.659$ & $0.293$ \\ \smallskip 
1.399--1.418 & ${0.12118}_{-0.00035}^{+0.00037}$ &  $0.549$ & $0.293$ & $-0.634$ & $0.286$ \\ \smallskip 
1.418--1.437 & ${0.12156}_{-0.00036}^{+0.00035}$ &  $0.567$ & $0.248$ & $-0.601$ & $0.278$ \\ \smallskip 
1.437--1.455 & ${0.12164}_{-0.00036}^{+0.00035}$ &  $0.590$ & $0.180$ & $-0.534$ & $0.252$ \\ \smallskip 
1.455--1.474 & ${0.12167}_{-0.00036}^{+0.00035}$ &  $0.608$ & $0.154$ & $-0.521$ & $0.249$ \\ \smallskip 
1.474--1.492 & ${0.12088}_{-0.00040}^{+0.00039}$ &  $0.622$ & $0.107$ & $-0.474$ & $0.232$ \\ \smallskip 
1.492--1.511 & ${0.12160}_{-0.00037}^{+0.00037}$ &  $0.634$ & $0.082$ & $-0.459$ & $0.229$ \\ \smallskip 
1.511--1.529 & ${0.12101}_{-0.00039}^{+0.00039}$ &  $0.667$ & $-0.002$ & $-0.388$ & $0.209$ \\ \smallskip 
1.529--1.548 & ${0.12053}_{-0.00039}^{+0.00042}$ &  $0.690$ & $-0.058$ & $-0.352$ & $0.202$ \\ \smallskip 
1.548--1.567 & ${0.12123}_{-0.00042}^{+0.00044}$ &  $0.715$ & $-0.098$ & $-0.334$ & $0.199$ \\ \smallskip 
1.567--1.585 & ${0.12057}_{-0.00041}^{+0.00041}$ &  $0.723$ & $-0.147$ & $-0.276$ & $0.176$ \\ \smallskip 
1.585--1.604 & ${0.12030}_{-0.00041}^{+0.00042}$ &  $0.723$ & $-0.170$ & $-0.241$ & $0.161$ \\ \smallskip 
1.604--1.622 & ${0.12071}_{-0.00041}^{+0.00042}$ &  $0.751$ & $-0.227$ & $-0.201$ & $0.150$ \\ \smallskip 
1.622--1.641 & ${0.11954}_{-0.00043}^{+0.00042}$ &  $0.754$ & $-0.214$ & $-0.214$ & $0.154$ \\ \hline 
\end{tabular}

\end{minipage}
\end{table}

In addition to the spectroscopic WFC3 data, we fit the ground-based photometric transit data presented in \cite{2016MNRAS.tmp..312D}, comprised of three $B$ band lightcurves, two $r^\prime$ lightcurves, and four $z^\prime$ lightcurves. For all ground-based lightcurves we used a GP systematics model, with a transit signal multiplied by a linear time trend as the mean function. For the covariance matrix, we adopted a Mat\'{e}rn $\nu=3/2$ kernel with multiple input variables such as the $xy$ coordinates of the star on the detector, seeing, and airmass. Lightcurves in the same bandpass were fit simultaneously, with $\RpRs$ shared but $\Tmid$, linear trends, covariance parameters, and white noise levels allowed to vary separately. Nonlinear limb darkening laws were adopted, with coefficients fixed to the values reported in Table \ref{table:spectroscopic_lightcurves}, which were taken from the catalogues of \cite{2000A&A...363.1081C,2003A&A...401..657C}. Remaining transit parameters were fixed to the values adopted for the WFC3 analysis, namely, those reported by \cite{2016MNRAS.tmp..312D} and listed in Table \ref{table:white_lightcurve}. Fitting was performed by locating the maximum likelihood and marginalizing over the parameter space with affine-invariant MCMC, as described above. Results are reported in Table \ref{table:spectroscopic_lightcurves}.

\section{Transmission spectrum} \label{sec:transmission_spectrum}

\begin{figure*}
\centering  
\includegraphics[width=\linewidth]{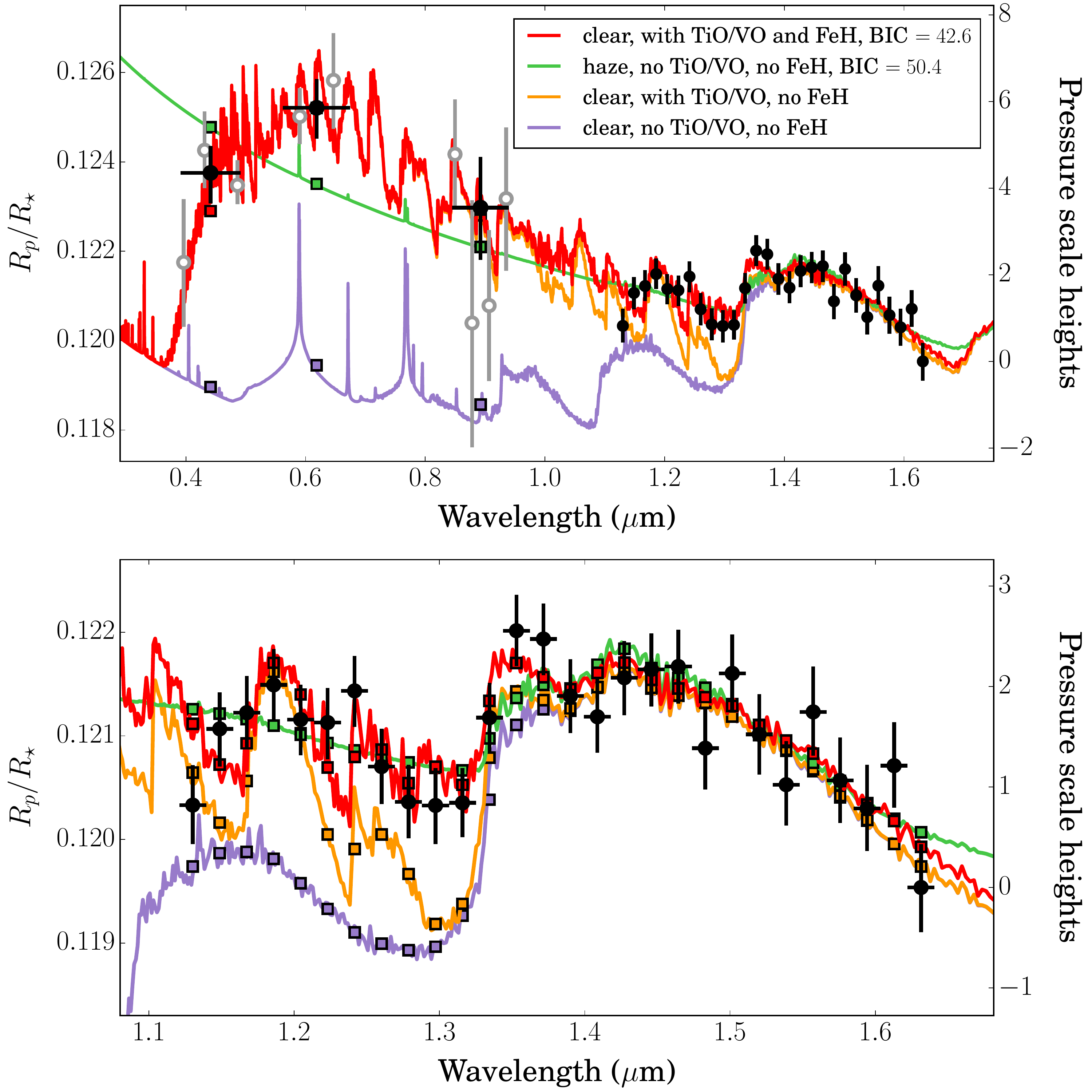}
\caption{Measured transmission spectrum for WASP-121b, with models. In the top panel, unfilled gray data points show results for fits to individual photometric lightcurves, with horizontal offsets applied for clarity, and filled black data points show results for joint fits to all lightcurves in the corresponding bandpass. The lower panel shows a zoomed-in view of the WFC3 transmission measurements. Models assume solar abundances and chemical equilibrium, with the exception of TiO, VO, and FeH, the abundances of which were adjusted to fit the data. (Red line) Clear atmosphere including TiO, VO, and FeH opacity. (Orange line) Same as previous, but excluding FeH opacity to illustrate its effect. (Green line) Hazy atmosphere with enhanced Rayleigh scattering, excluding TiO, VO, and FeH opacity. (Purple line) Clear atmosphere excluding TiO, VO, and FeH opacity.}
\label{fig:transmission_spectrum}
\end{figure*}

The measured transmission spectrum for WASP-121b is plotted in Figure \ref{fig:transmission_spectrum}. The WFC3 data show structure, with two broad features centered around $1.2\,\um$ and $1.4\,\um$. The scale of these variations is on the order of a few atmospheric scale heights, consistent with expectations for molecular absorption. Additionally, the effective planetary radius is found to be significantly larger at optical wavelengths relative to  near-infrared wavelengths. 

Before exploring implications for the planetary atmosphere, we considered the possibility of stellar activity causing the different radii measured at optical and near-infrared wavelengths. Making the conservative assumption that star spots are non-luminous, we calculate that the unocculted spot coverage would need to decrease by an amount equivalent to 7\% of the visible stellar disc between the optical and near-infrared observation epochs. This seems highly unlikely, especially given that WASP-121 appears photometrically stable at the millimag level \citep{2016MNRAS.tmp..312D}. Also, analyses of the individual lightcurves for each bandpass produced consistent results at the $1\sigma$ level (top panel of Figure \ref{fig:transmission_spectrum}), suggesting that stellar activity does not affect the inferred radius significantly from epoch to epoch. We therefore conclude that variations in the inferred radius shown in Figure \ref{fig:transmission_spectrum} are due to the planetary atmosphere.

We used the 1D radiative transfer code \texttt{ATMO} to generate model spectra and investigate atmospheric opacity sources \citep{2014A&A...564A..59A,2015ApJ...804L..17T,2016ApJ...817L..19T}. Isothermal temperature-pressure profiles were assumed, with temperature allowed to vary in the fitting. We produced models with and without TiO, VO, and FeH. The latter species are observed in M/L dwarfs \citep{2001RvMP...73..719B} and models predict they may also be present in hot gas giant planets \citep[e.g.][]{2007ApJS..168..140S}. For models including TiO/VO/FeH, we varied the relative abundances of TiO/H$_2$O, VO/H$_2$O, and Fe/H$_2$O during fitting. Solar abundances under thermochemical equilibrium were adopted for other gas species, including H$_2$O. We note that for transmission spectra, the absolute abundances of individual species are often unconstrained owing to a well-known degeneracy with the absolute pressure level, but the relative abundances between species can be well measured \citep{2008A&A...481L..83L}.  In our fitting, TiO/VO/FeH abundances were adjusted using the analytical relation of \cite{2008A&A...481L..83L} for the wavelength-dependent transit radius of an isothermal atmosphere. We also considered a simple haze scattering model in the form of a Rayleigh profile with collisional cross-section area allowed to vary as a free parameter, as described in \cite{2016Natur.529...59S}. 

Best fits for a number of illustrative models are shown in Figure \ref{fig:transmission_spectrum}, obtained using the nonlinear least squares IDL routine \texttt{mpfit} \citep{2009ASPC..411..251M}. We find that a model excluding TiO, VO, FeH, and haze is unable to account for the data (purple line). In particular, it fails to reproduce the optical photometry and short wavelength WFC3 data. A good fit is achieved, however, when TiO, VO, and FeH opacity is included (red line). For this model, we find a temperature of $1500 \pm 230$\,K, and abundances relative to thermochemical equilibrium of $\sim 7\times$ solar for TiO/H$_2$O, $\sim 5 \times$ solar for VO/H$_2$O, and $\sim 0.2 \times$ solar for FeH/H$_2$O. The inferred temperature is lower than the dayside equilibrium temperature of $\Teq \sim 2400$\,K, which may indicate a significantly cooler nightside hemisphere. The strong absorption of TiO and VO in the optical accounts for the larger radii measured at these wavelengths relative to the near-infrared. To illustrate the effect of FeH, we also generated a model with it removed, but TiO and VO retained (orange line). Such a model is unable to reproduce the WFC3 transmission spectrum at wavelengths near $1.3\um$, where FeH has prominent absorption. Lastly, a model excluding TiO, VO, and FeH, but including haze, gives a relatively poor fit to the data (green line). Most significantly, haze scattering underpredicts the $r^{\prime}$ opacity by $2.6\sigma$. To assess the relative quality of fit between the haze and TiO/VO/FeH models, we computed the Bayesian information criterion (BIC) for each, obtaining $\textnormal{BIC}=50.4$ for the haze model and $\textnormal{BIC}=42.6$ for the TiO/VO/FeH model. Following \cite{kassr95}, we approximate the Bayes factor as $\exp(-\Delta \textnormal{BIC}/2) = 0.02$ in favor of the TiO/VO/FeH model over the haze model. This strongly supports the hypothesis that TiO and VO, rather than haze, are responsible for the enhanced opacity measured at optical wavelengths. 

All models favor H$_2$O absorption, which matches the measured spectrum well between $1.3$--$1.64\um$. To quantify the significance of this detection, we compared the quality of fit obtained for the best-fit TiO/VO/FeH model described above with that obtained for the same model but H$_2$O removed. We determine that H$_2$O is detected at a confidence level of $\watersig\sigma$. 

\section{Discussion} \label{sec:discussion}

Our observations reveal H$_2$O absorption in the atmosphere of WASP-121b with an amplitude of $>2$ gas pressure scale heights. This result is consistent with a clear atmosphere, assuming TiO/VO and FeH absorption is responsible for the additional opacity measured across the 1.12--1.3$\um$ wavelength range. In contrast, WASP-12b is observed to have a hazy atmosphere without TiO/VO \citep{2013MNRAS.436.2956S}, despite having similar properties to WASP-121b, reinforcing the emerging picture of hot gas giant diversity \citep{2016Natur.529...59S}. We note, however, that the optical data cannot currently exclude a model including both TiO/VO absorption and haze scattering for WASP-121b. Such a model would be very similar to our best-fit model (red line in Figure \ref{fig:transmission_spectrum}), except that the effective radius would continue increasing for wavelengths shortward of the $B$ bandpass. To confidently discount such a scenario, it will be necessary to spectrally resolve the characteristic steep rise in opacity at $\sim 0.35\um$ due to TiO/VO with further observations. 

Until then, the evidence for TiO/VO absorption in the atmosphere of WASP-121b remains tentative, as it primarily hinges upon the relative radius measured from the ground in the $r^{\prime}$ bandpass. We emphasize, however, that this measurement is based on two lightcurves that produce consistent results when analyzed separately and simultaneously (top panel of Figure \ref{fig:transmission_spectrum}). In addition,  \cite{2016MNRAS.tmp..312D} report ground-based secondary eclipse measurements in the $z^{\prime}$ bandpass that imply a brightness temperature higher than the equilibrium temperature predicted for zero Bond albedo and instantaneous heat re-radiation. This could potentially be explained by a thermal inversion, with TiO being observed in emission in the $z^{\prime}$ bandpass. Again, follow-up observations that spectrally resolve the TiO/VO features will be necessary for confirmation.

To date, \textit{HST} spectroscopy data have been published for three exoplanets with $\Teq > 2000$\,K. Of these, TiO/VO absorption has been ruled out for WASP-12b \citep[$\Teq \sim 2600$\,K;][]{2013MNRAS.436.2956S} and WASP-19b \citep[$\Teq\sim 2100$\,K;][]{2013MNRAS.434.3252H}, while evidence for TiO emission has been reported for WASP-33b \citep[$\Teq \sim 2700$\,K;][]{2015ApJ...806..146H}. It is possible that WASP-33b is hot enough to maintain TiO/VO in the gas phase throughout its dayside atmosphere, whereas the day-night terminators of WASP-12b and WASP-19b are too cool, or cold-trapping occurs elsewhere in their atmospheres. However, the fact that WASP-121b has a lower equilibrium temperature ($\Teq \sim 2400$\,K) than WASP-12b, and yet shows evidence for TiO/VO absorption at low pressures, indicates that additional factors are at play. For instance, models predict larger temperature contrasts between the dayside and nightside hemispheres with increasing stellar irradiation and planet metallicity \citep{2008ApJ...678.1419F,2008ApJ...673..513D,2015ApJ...801...86K}. It is thus plausible that WASP-12b, with its higher irradiation and host star metallicity, has a cooler nightside than WASP-121b, which could result in a nightside cold-trap for WASP-12b but not WASP-121b. Compositional differences could also have a significant influence on the temperature profiles for each planet \citep{2014A&A...562A.133P,2016ApJ...817L..19T}, in turn determining the susceptibility to cold-trapping.

\section{Conclusion} \label{sec:conclusion}

We have presented a transmission spectrum for WASP-121b spanning the $1.12$--$1.64\um$ wavelength range. Absorption by H$_2$O is detected at high confidence ($\watersig\sigma$), consistent with predictions of clear atmosphere models. Deeper transits measured at optical wavelengths relative to the near-infrared strongly favor models including TiO/VO absorption, but scattering by a high-altitude haze cannot yet be definitively excluded. We also find evidence for FeH absorption in the WFC3 bandpass. WASP-121b is one of the most favorable targets available for both transmission and emission spectroscopy, and offers particular promise for exploring the link between strong optical absorbers, such as TiO/VO, and thermal inversions in hot gas giant atmospheres. 

\acknowledgements{Based on observations made with the NASA/ESA Hubble Space Telescope, obtained at the Space Telescope Science Institute, which is operated by the Association of Universities for Research in Astronomy, Inc., under NASA contract NAS 5-26555. The authors are grateful to the WASP-121 discovery team for generously providing the ground-based photometric lightcurves. Support for this work was provided by NASA through grants under the HST-GO-14468 program from the STSci. The research leading to these results has received funding from the European Research Council under the European Union Seventh Framework Program (FP7/2007-2013) ERC grant agreement no.\ 336792. HRW acknowledges support by an appointment to the NASA Postdoctoral Program at Goddard Space Flight Center, administered by ORAU and USRA through a contract with NASA. NPG gratefully acknowledges support from the Royal Society in the form of a University Research Fellowship.}

\end{document}